\documentclass[prapplied, reprint, superscriptaddress, floatfix, longbibliography]{revtex4-2}

\usepackage{mathtools}
\usepackage{textgreek}
\usepackage{graphicx}
\usepackage{bm}
\usepackage{hyperref}
\usepackage{braket}
\usepackage{xcolor}
\usepackage{multirow}
\usepackage{placeins}
\usepackage{mleftright}
\usepackage[norefs, nocites, ignoreunlbld]{refcheck}
\usepackage{physics}
\usepackage{subcaption}
\usepackage{booktabs}

\usepackage{amssymb}
\usepackage{ragged2e}
\usepackage[justification=justified, singlelinecheck=false, format=plain]{caption}
\usepackage{tabularx}
\usepackage{float}

\begin{document}

\title{Temporal-Mode Engineering for Multiplexed Microwave Photons\\ and Mode-Selective Quantum State Transfer}

\author{Keika~Sunada$^{\dag}$}
\thanks{These authors contributed equally:\\
$\dag$ k.sunada@qipe.t.u-tokyo.ac.jp\\
$\ddag$ miyamura@qipe.t.u-tokyo.ac.jp}

\author{Takeaki~Miyamura$^{\ddag}$}
\thanks{These authors contributed equally:\\
$\dag$ k.sunada@qipe.t.u-tokyo.ac.jp\\
$\ddag$ miyamura@qipe.t.u-tokyo.ac.jp}
\author{Kohei~Matsuura}
\affiliation{Department of Applied Physics, Graduate School of Engineering, The University of Tokyo, Bunkyo-ku, Tokyo 113-8656, Japan}
\author{Zhiling~Wang}
\affiliation{RIKEN Center for Quantum Computing (RQC), Wako, Saitama 351-0198, Japan}
\author{Jesper~Ilves}
\affiliation{Department of Applied Physics, Graduate School of Engineering, The University of Tokyo, Bunkyo-ku, Tokyo 113-8656, Japan}
\author{Shingo~Kono}
\affiliation{
NNF Quantum Computing Programme, Niels Bohr Institute, University of Copenhagen, Denmark}
\author{Yasunobu~Nakamura}
\affiliation{Department of Applied Physics, Graduate School of Engineering, The University of Tokyo, Bunkyo-ku, Tokyo 113-8656, Japan}
\affiliation{RIKEN Center for Quantum Computing (RQC), Wako, Saitama 351-0198, Japan}

\date{\today}

\begin{abstract}
Quantum communication between distant superconducting qubits on separate chips using itinerant microwave photons has been studied to realize distributed quantum information processing. To enhance information capacity and fault tolerance in quantum networks, it is beneficial to encode a large quantity of quantum information using auxiliary degrees of freedom of these photons.
In this work, we experimentally investigate the use of temporal modes of photon wave packets.
Through the photon-shaping technique with a fixed-frequency transmon qubit, we generate single microwave photons in four orthogonal temporal modes propagating along a waveguide. We demonstrate mode-selective absorption across orthogonal modes via the time-reversed process of emission, achieving absorption efficiencies exceeding $0.89$ for mode-matched cases, while remaining below $0.13$ for orthogonal modes.
Photons rejected by a given receiver mode can remain mutually orthogonal, enabling selective absorption at subsequent receivers in future multi-node architectures.
These results highlight the feasibility of temporal-mode engineering for constructing a higher-dimensional orthogonal basis for multiplexed quantum networks.
\end{abstract}

\maketitle

\section{INTRODUCTION}
Large-scale quantum computing with superconducting circuits will require integration of numerous qubits to perform real-world applications~\cite{fowler_surface_2012}. As the number of qubits on a single chip increases, challenges related to the chip-size limitation~\cite{van_damme_advanced_2024}, yield rate~\cite{hertzberg_laser-annealing_2021, osman_mitigation_2023}, and heat load of wirings~\cite{SKrinner_heatload_2019} become more pronounced. Distributed quantum computing~\cite{ang_architectures_2022,caleffi_distributed_2024}, in which multiple quantum processors work cooperatively, has emerged as a promising approach to address these scalability limitations. This architecture necessitates quantum channels capable of faithfully transferring quantum states between spatially separated processing units.
In superconducting quantum systems, propagating microwave photons provide a natural means for establishing quantum communication channels between remote qubits. Several experimental demonstrations have successfully achieved deterministic quantum state transfer and remote entanglement between superconducting qubits separated by distances up to tens of meters~\cite{kurpiers_deterministic_2018, campagne-ibarcq_deterministic_2018, axline_-demand_2018,Leung_deterministic_2019, storz_loophole-free_2023, grebel_bidirectional_2024, qiu_deterministic_2025, almanakly_deterministic_2025, TMiyamura_qcomm_2025}.

Leveraging additional degrees of freedom of photons offers multiple advantages for scalable quantum networks. First, it can help mitigate the effects of photon loss during transmission through dual-rail encoding. Various encoding schemes have been explored for this purpose and have been experimentally demonstrated. Time-bin encoding employs temporally separated photonic modes~\cite{kurpiers_quantum_2019, ilves_-demand_2020}, while frequency-bin encoding utilizes distinct frequency modes~\cite{yang_deterministic_2024, JOsullivan_frequebcy-bin_2025, ZWang_freq-bin_2025}. Path encoding uses spatially separated transmission channels~\cite{kurpiers_quantum_2019,kannan_waveguide_2020, kannan_generating_2020}. 
Moreover, these schemes increase the overall information capacity of quantum channels and enable multiplexed communication~\cite{penas_multiplexed_2024}. However, implementing higher-dimensional encoding with each of these approaches involves trade-offs in temporal window and frequency bandwidth.

In the optical domain, temporal modes, defined by orthogonal temporal profiles of propagating photons, have attracted attention for their high distinguishability and accessibility to high-dimensional Hilbert spaces~\cite{Raymer_TM_review_2020}.
Selective control over these modes has been experimentally demonstrated \cite{Reddy_QPG_2018}, highlighting their potential as a basis for quantum information processing.
In the microwave domain, temporal-mode multiplexing has been theoretically proposed~\cite{penas_multiplexed_2024}.

In this work, we experimentally demonstrate temporal-mode engineering through controlled emission of itinerant microwave photons and their selective absorption. Using superconducting circuits with fixed-frequency transmons and a photon-shaping technique~\cite{miyamura_generation_2025}, we generate single photons in four orthogonal temporal modes, achieving high mode orthogonality. We further observe selective absorption based on their temporal-mode matching: the absorption efficiencies exceed $0.89$ for matched modes, while remaining below $0.13$ for orthogonal modes.
In addition, we discuss the preservation of the orthogonality among the rejected photons, as well as the preservation of their quantum states, during the mode-selective absorption process.
Our results highlight that the temporal modes constitute a highly orthogonal basis, suggesting that this scheme offers a feasible approach to expanding the Hilbert-space dimension of quantum network channels.

\section{Theory of Temporal-Mode Multiplexing}
\begin{figure}[t]
    \centering
    \includegraphics{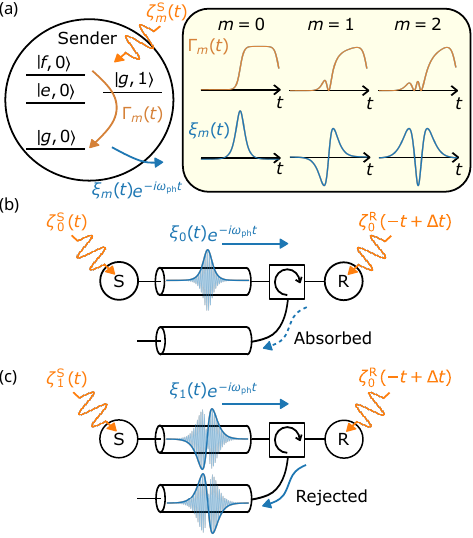}
    \caption{\justifying Basic concepts of temporal-mode multiplexing and selective absorption of microwave photons using a superconducting qubit. 
    (a)~Temporal-mode engineering via a resonator-assisted Raman process. The photon mode $m$ with the waveform $\xi_m(t)$, is engineered through the effective decay rate $\Gamma_m(t)$, controlled by the sender's drive $\zeta^\text{S}_m(t)$. 
    (b)~Mode-matched photon absorption. A photon in mode $0$ with the waveform $\xi_0(t)e^{-i\omega_\text{ph}t}$, generated by the sender under the drive $\zeta^\text{S}_0(t)$, is perfectly absorbed by the receiver with the drive $\zeta^{\text{R}}_0(-t+\Delta t)$. 
    (c)~Orthogonal-mode photon rejection. A photon in mode $1$ with the mode function $\xi_1(t)e^{-i\omega_\text{ph}t}$ generated by the sender under the drive $\zeta^\text{S}_1(t)$, is perfectly rejected by the receiver with the drive $\zeta^{\text{R}}_0(-t+\Delta t)$.}
    \label{fig:1}
\end{figure}

\subsection{Orthogonal temporal modes of itinerant photons}
The temporal profile of a photon is defined by its mode function $\xi(t)$ with the normalization condition $\int_{-\infty}^{\infty} |\xi(t)|^2 dt = 1$.
By leveraging this temporal degree of freedom of photons, a set of orthonormal mode functions $\{\xi_m(t)\}$ can be constructed, where $m = 0, 1, \ldots$ labels the photon modes. 
The mutual overlap between different mode functions is defined as
\begin{equation}
    I_{mm'} \coloneqq \ev{\xi_m, \xi_{m'}} =  \int_{-\infty}^{\infty}\xi_m^\ast(t) \xi_{m'}(t)\,dt ,
\end{equation}
and the orthogonality condition is satisfied when
\begin{equation}
    I_{mm'} = \delta_{mm'},
    \label{eq:orthogonality_condition}
\end{equation}
where $\delta_{mm'}$ denotes the Kronecker delta, which is equal to 1 for $m=m'$ and 0 otherwise.
A set of orthogonal mode functions can be constructed via the Gram--Schmidt orthonormalization procedure~\cite{penas_multiplexed_2024}. 
Analytical expressions and plots of these mode functions, constructed in a real space, are provided in Appendix~\ref{sec:app2}, where the construction starts from a hyperbolic secant pulse, a pulse shape commonly used for itinerant photons~\cite{PMagnard_2020, kurpiers_deterministic_2018, axline_-demand_2018, campagne-ibarcq_deterministic_2018, miyamura_generation_2025, TMiyamura_qcomm_2025}.
Furthermore, the Fourier-transformed  mode function $\xi_m(\omega)$ also satisfies the orthogonality condition $\int_{-\infty}^{\infty}\xi_m^\ast(\omega) \xi_{m'}(\omega)\,d\omega =\delta_{mm'}$.

An itinerant photon propagating in a transmission line with mode function $\xi_m(t)$ can be described with creation and annihilation operators defined as a superposition of the field operators at time $t$, $\hat{a}_t$, and $\hat{a}_t^\dagger$:
\begin{align}
    \hat{a}_m &= \int_{-\infty}^{\infty} \xi^\ast_m(t)\, \hat{a}_t\, dt,\\
    \hat{a}_m^\dagger &= \int_{-\infty}^{\infty} \xi_m(t)\, \hat{a}_t^\dagger\, dt.
\end{align}
Here, the field operators satisfy the commutation relation $[\hat{a}_t,\hat{a}_{t'}^\dagger]=\delta(t-t')$.
The pulse mode operators corresponding to mode $m$ and $m'$ satisfy
\begin{equation}
    [\hat{a}_m, \hat{a}^\dagger_{m'}]  =  \int_{-\infty}^{\infty}\xi^\ast_m(t) \xi_{m'}(t)\,dt = \delta_{mm'}.
    \label{eq:commutation_relation}
\end{equation}
This indicates that pulse mode operators associated with orthogonal mode functions are commutative.
As a result, the modes behave as independent quantum channels that can be individually controlled and utilized.

\subsection{Temporal-mode engineering and selective absorption}
Temporal-mode engineering can be achieved using a transmon qubit~\cite{koch_charge-insensitive_2007} dispersively coupled to a resonator.
We denote the eigenstates of each system as $\ket{s, n}$, where $\ket{s}$ represents the qubit state ($\ket{g}$, $\ket{e}$, $\ket{f}$, corresponding to the three lowest energy levels), and $\ket{n}$ represents the resonator Fock state.
As illustrated in Fig.~\ref{fig:1}(a), photon emission is realized via a resonator-assisted Raman process~\cite{pechal_microwave-controlled_2014,zeytinoglu_microwave-induced_2015}, where a microwave drive at the sender induces the $|f,0\rangle$$\rightarrow$$|g,1\rangle$ transition.
Controlling this $|f\rangle$$\rightarrow$$|g\rangle$ qubit relaxation process accompanied with a photon emission enables photon waveform engineering.
To generate a photon in mode $m$, the drive-induced relaxation rate must satisfy~\cite{miyamura_generation_2025, mcintyre_protocols_2025}
\begin{equation}
    \Gamma_m(t) = \frac{|\xi_m(t)|^2}{1-\int^t|\xi_m(t')|^2\, dt'}.
    \label{eq:2B_Gamma_f}
\end{equation}
The rates $\Gamma_m(t)$ corresponding to the first few temporal modes $\xi_m(t)$ are shown in Appendix~\ref{sec:app2} and are implemented through the sender’s drive $\zeta^\text{S}_m(t)$.

Photon absorption is realized by implementing the corresponding time-reversed qubit excitation rate $\Gamma_m(-t+\Delta t)$ at the receiver~\cite{JICirac_commtheory_1997}, which is an equivalent system to the sender.
The temporal offset $\Delta t$ is optimized by shifting the receiver drive $\zeta^{\text{R}}_m(-t+\Delta t)$ to achieve temporal mode matching, effectively reversing the emission process in mode $m$.
The drive modes of the sender, $m$, and of the receiver, $n$, can be independently configured.  
They define the temporal profiles of the emitted and absorbable photon modes, respectively.  
Such independent controllability enables selective photon absorption based on the mode orthogonality, as indicated in Eq.~\eqref{eq:commutation_relation}.
In the ideal case, a photon emitted in mode $m$ is perfectly absorbed by the receiver drive mode $n$ when $m=n$, as exemplified in Fig.~\ref{fig:1}(b), and rejected when $m\neq n$, as in Fig.~\ref{fig:1}(c).

\section{EXPERIMENT}
We prepare two devices with a fixed-frequency transmon, a sender and a receiver, with similar parameters, connected via coaxial cables and routed through a circulator. 
The photon flux reflected at the receiver is amplified in a phase-insensitive manner with a flux-driven Josephson parametric amplifier~(JPA)~\cite{yamamoto_JPA_2008} and directed to the output line.
Details of the experimental setup are provided in Appendix~\ref{sec:app1}.

\subsection{Photon generation in orthogonal temporal modes}
\begin{figure}
    \centering
    \includegraphics{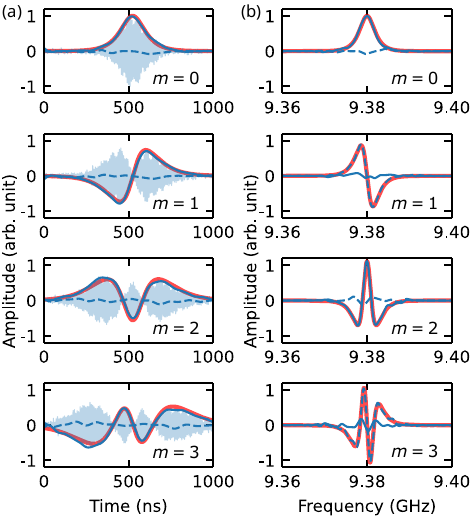}\\
    \includegraphics{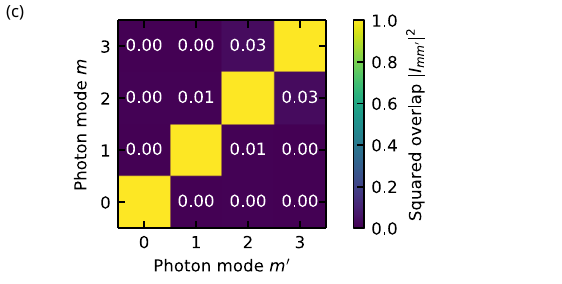}
    \caption{\justifying Photon generation in orthogonal temporal modes $m=0,\,\ldots,\,3$. 
    (a)~Measured demodulated photon modes in the time domain.
    Light blue solid lines show the averaged photon waveforms emitted from the sender.
    Dark blue lines show the real (solid) and imaginary (dashed) parts of the modes $\xi_m^\text{S}(t)$.
    Coral solid lines represent the real parts of the target modes.
    (b)~Measured demodulated photon modes in the frequency domain.
    Dark blue lines show the real (solid) and imaginary (dashed) parts of the Fourier amplitude of the modes $\xi_m^\text{S}(\omega)$.
    Coral solid lines indicate the target spectra: the real part for even modes ($m=0$ and 2) and the imaginary part for odd modes ($m=1$ and 3).
    (c)~Squared-overlap matrix $\{|I_{mm'}|^2\}$ obtained from the experimentally generated waveforms. }
    \label{fig:2}
\end{figure}

Following the method demonstrated in Ref.~\citenum{miyamura_generation_2025}, we determine the sender's drive $\zeta^\text{S}_m(t)$ required to implement the desired time-dependent decay rate $\Gamma_m(t)$.
As a result, we obtain the target photon mode with the waveform $\xi_m(t)$ at a fixed carrier frequency $\omega_\text{ph}$.
Note that the sign change in $\xi_m(t)$ necessary for orthogonality is achieved through the control of the drive phase, as the photon directly inherits this phase.

In the following experiments, we set the carrier frequency of the emitted photon $\omega_\text{ph}/2\pi$ to $9.38$ GHz, which lies within the bandwidths of both the sender and the receiver, and the pulse-width parameter $\kappa_\text{ph}/2\pi$ [See for the definition Eq.~\eqref{eq:kappa_ph_def} in Appendix~\ref{sec:app2}.] to $5$ MHz, which is experimentally achievable.
The initial sender state is prepared as $|\psi\rangle_{\pm}^\mathrm{S}=(\ket{g}\pm\ket{e})/\sqrt{2}$.
Then, a $\pi_{ef}$ pulse followed by the sender's drive $\zeta^\text{S}_m(t)$ induces the transition between $\ket{f, 0}$ and $\ket{g, 1}$, resulting in a photon emission into the transmission line.
To suppress background noise without distorting the photon waveform, we subtract the measurement results obtained from the initial state $|\psi\rangle_{-}^{\mathrm{S}}$ from those obtained from $|\psi\rangle_{+}^{\mathrm{S}}$.
We additionally alternate the phase of the pumping pulse and combine the corresponding results, allowing us to amplify the signal in a phase-insensitive manner~\cite{YSunada_phase-insensitive_2024}.
Here, we generate photon modes indexed by $m=0,\,\ldots,\,3$. The observed photon waveforms $\xi_m^\text{S}(t)e^{-i\omega_\text{ph}t}$ emitted from the sender are shown in Fig.~\ref{fig:2}(a).
Their Fourier spectra $\xi_m^\text{S}(\omega)$ are presented in Fig.~\ref{fig:2}(b).
Although these photons are reflected by the receiver before detection, the distortion is negligible because of the large linewidth of the receiver resonator.

To quantify how well the generated photon waveforms $\xi_m^\text{S}(t)$ form an orthogonal basis, we evaluate their squared-overlap matrix:
\begin{equation}
    \left\{ |I_{mm'}|^2 \right\}=\left\{\, | \langle \xi_m^\text{S}(t), \xi_{m'}^\text{S}(t)\rangle |^2 \,\right\},
\end{equation}
as shown in Fig.~\ref{fig:2}(c).
The overlaps between modes are largely suppressed, although small residual overlaps remain, particularly for higher-order modes. 
In Fig.~\ref{fig:2}(a), the observed waveforms exhibit small but noticeable imaginary components, even though the target mode functions are defined as purely real-valued in the time domain. 
Similarly, in Fig.~\ref{fig:2}(b), the spectra show minor residual imaginary (real) components, which ideally should vanish for even (odd) modes. 
These phase distortions are largely compensated for, but not completely eliminated.

\subsection{Mode-selective absorption}
\begin{figure}
    \centering
    \includegraphics{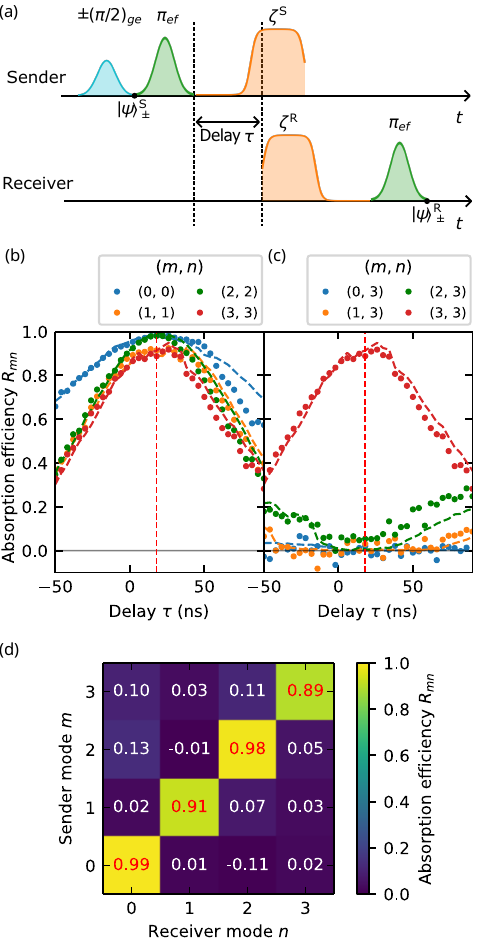}
    \caption{\justifying Mode-selective absorption.
    (a)~Pulse sequence for the absorption process. The initial sender state $\ket{\psi}_\pm^\text{S}=(\ket{g}\pm\ket{e})/\sqrt{2}$ is prepared by a $\pm(\pi/2)_\text{ge}$ pulse. Then, a $\pi_\text{ef}$ pulse followed by the sender's drive $\zeta^\text{S}(t)$ induces the $|f,0\rangle$--$|g,1\rangle$ transition. 
    After a time delay $\tau$, a time-reversed receiver's drive $\zeta^{\text{R}}(-t+\Delta t)$ and a $\pi_\text{ef}$ pulse are applied to the receiver.
    (b)~Mode-matched photon absorption as a function of the delay $\tau$. 
    (c)~Mode-selective absorption.
    In~(b) and~(c), dots shows the absorption efficiency $R_{mn}$ as a function of the delay $\tau$ of the drive pulse. Dashed lines show the squared-overlap of the emitted and absorbed photon modes, $|I'_{mn}|^2$.
    (d)~Summary of the mode-selective absorption efficiencies $R_{mn}$ with the optimal delay time $\tau^\mathrm{opt}$~[vertical dashed lines in~(b) and~(c)].}
    \label{fig:3}
\end{figure}

To demonstrate the selective absorption of multiplexed photons in temporal modes, we independently control the sender mode $m$ and the receiver mode $n$.
The receiver's drive $\zeta_n^\text{R}(t)$ is calibrated through the same procedure as for the sender, targeting the photon waveforms $\xi_n(t)$ as used in the sender calibration.
We observe the emitted-photon modes from the receiver, which we denote as $\xi_n^\text{R}(t)$.
We perform photon absorption measurement by applying the time-reversed drive $\zeta_n^{\text{R}}(-t+\Delta t)$ and sweeping the temporal offset $\Delta t$ by varying the relative delay $\tau$ between the sender's and receiver's drive pulses.
After applying a $\pi_\text{ef}$ pulse at the receiver, perfect state transfer from the sender to the receiver would result in $\ket{\psi}^\text{R} = \ket{\psi}^\text{S}$.
The sequence of this operation is illustrated in Fig.~\ref{fig:3}(a).
The absorption efficiency is defined as~\cite{kurpiers_deterministic_2018}
\begin{equation}
    R_{mn} = 1 - \frac{E^\text{res}_{mn}}{E^\text{base}_m},
\end{equation}
where $E^\text{res}_{mn}=\int|\xi_{mn}^\text{S, res}(t)|^2\,dt$ represents the residual photon flux under the mode-$m$ emission and mode-$n$ absorption processes, while $E^\text{base}_m=\int| \xi_{m}^\text{S}(t)|^2\,dt$ is the baseline photon flux under the mode-$m$ emission and no absorption process (with the receiver deactivated).

When the modes are matched ($m = n$), the absorption efficiency $R_{mm}$ ($m = 0,\,\ldots,\,3$) shows a peak at the optimal delay, as shown in Fig.~\ref{fig:3}(b).
Here, we define the optimal delay $\tau^\text{opt}$ as the time corresponding to the maximum absorption for mode $m=0$, which is $\tau^\text{opt} = 18$~ns.
This value reflects not only the physical photon propagation time but also an additional effective shift arising from the truncation of the sender’s drive pulse~\cite{PMagnard_2020}.
We also examine the absorption efficiencies $R_{m3}$ ($m = 0,\,\ldots,\,3$), where we fix the receiver mode to $n=3$, as shown in Fig.~\ref{fig:3}(c).
At the optimal timing, $R_{33}$ exhibits high absorption efficiency, while $R_{m3}$ for $m = 0,\,1,$ and~2 is suppressed.
The difference in $R_{mn}$ depending on the selected mode demonstrates the successful selective absorption.

These results can be directly estimated from the temporal mode overlap between the photon emitted by the sender, $\xi_m^\text{S}(t)$, and the absorbable mode of the receiver, $\xi_n^{\text{R}}(-t+\Delta t)$:
\begin{align}
I'_{mn} = \langle \xi_m^\text{S}(t), \xi_n^{\text{R}}(-t+\Delta t)\rangle.
\label{eq:abs_estimation}
\end{align}
This expression follows from decomposing the emitted waveform into a component parallel to the absorbable mode and an orthogonal component that remains unabsorbed.
The delay-dependent squared-overlaps $|I'_{mn}|^2$, accounting for the delay $\tau^\text{opt}$, are plotted as dashed lines in Figs.~\ref{fig:3}(b) and (c), capturing the features of the experimental data points $R_{mn}$.
Here, the phase distortion of the emitted photon waveform $\xi_m^\mathrm{S}(t)$, originating from the instability of the sender’s drive over the long measurement duration, is independently characterized at each delay point and incorporated into the analysis.

A summary of the absorption efficiencies for all mode combinations is shown in Fig.~\ref{fig:3}(d).
Along the diagonal ($m = n$), where the selected modes match, we observe high absorption efficiencies ($R \geq 0.89$).
Conversely, the off-diagonal elements~($m\ne n$), corresponding to mutually orthogonal basis functions, exhibit relatively low efficiencies ($R \leq 0.13$) across the four modes.
For modes 1 and 3, reduced absorption efficiencies are observed even when the modes are matched ($m = n$), which is attributed to temporary phase distortions of the emitted photons.
We also observe some data points of $R_{mn}$ show negative values, which may arise from gain instabilities of the JPA.

\section{DISCUSSION}
We have investigated the feasibility of using temporal modes as a degree of freedom of itinerant microwave photons for multiplexing in quantum communication channels.
First, we generated single photons in four orthogonal temporal modes with high orthogonality, realized through the precise temporal control of the sender's drive.
We then demonstrated mode-selective absorption using these photons in orthogonal temporal modes.
By independently selecting the drive modes of the sender and receiver, we observed efficient absorption for mode-matched photons ($R \geq 0.89$) and rejection for orthogonal ones ($R \leq 0.13$).
These results highlight the potential of temporal-mode encoding as a feasible approach to enhance multiplexing capability in quantum communication channels.
Improved stability of the drive pulses and the JPA pump pulses is expected to further enhance the mode selectivity and the data reliability, respectively.
Based on the residual phase distortion in Fig.~\ref{fig:2}(a) after phase compensation, we estimate using Eq.~\eqref{eq:abs_estimation} that further suppression of time-dependent instabilities would allow the absorption efficiency to exceed 0.96 for mode-matched cases and to fall below 0.01 for orthogonal modes.
Such instabilities can arise from temperature fluctuations in the room-temperature amplifiers, mixers, and other components.
From an experimental perspective, important approaches include improving the long-term stability of the experimental setup, as well as increasing the pulse-width parameter $\kappa_\mathrm{ph}$, which shortens the total measurement duration.

\begin{figure}
    \centering
    \includegraphics{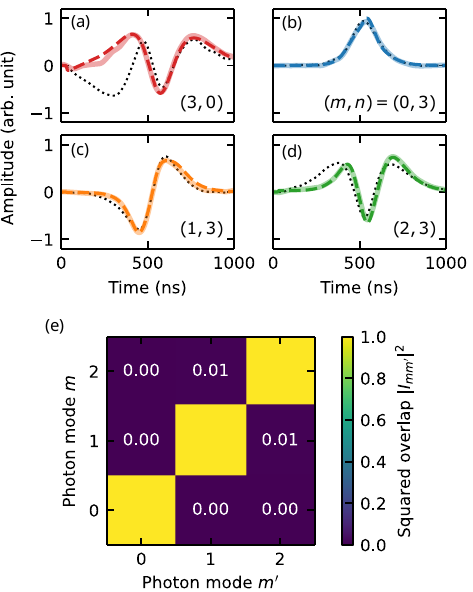}
    \caption{\justifying 
    Rejected photons during mode-selective absorption when the receiver mode does not match the incoming photon mode ($m\neq n$). 
    (a)~Rejected photon waveforms for $(m,n)=(3,0)$. 
    (b--d)~Rejected photon waveform for receiver drive fixed at $n=3$ and incoming photon modes $m=0,\,1,$ and~2. 
    In (a--d), the amplitude of the rejected photon waveforms $\xi_{mn}^\mathrm{S,res}(t)$ is shown by solid lines, with corresponding simulated results shown by dashed lines. Black dotted lines indicate the incoming photon waveforms $\xi_{m}^\mathrm{S}(t)$. 
    (e)~Squared-overlap matrix calculated from the experimentally measured rejected waveforms $\xi_{m3}^\mathrm{S,res}(t)$ for $m=0,\,1,$ and~2.
    }
    \label{fig:5}
\end{figure}
In future multiplexed quantum communication architectures, multiple sender and receiver nodes can be connected in cascade via circulators and coaxial cables. Photons rejected at the first receiver then propagate unidirectionally downstream and can be selectively absorbed by a subsequent receiver.
In this context, the rejected photons must preserve their mutual orthogonality.
Previous theoretical work~\cite{penas_multiplexed_2024}, however, has pointed out that waveform scattering during the rejection process generally leads to distortions in the outgoing photon waveforms.
In our experiment, such distortions are observed when the receiver mode index is smaller than that of the sender mode ($m>n$). In this case, the receiver cannot follow the rapid temporal variations of the incoming field $\xi_{m}^\text{S}(t)$, causing the first two temporal peaks to merge. As a result, the rejected photon flux $\xi_{mn}^\text{S,res}(t)$ exhibits a reduced number of peaks, as shown in Fig.~\ref{fig:5}(a) for the case $(m,n)=(3,0)$, together with the corresponding simulation results (see Appendix~\ref{sec:app4} for details).
In contrast, when $m<n$, the rejected waveforms largely preserve the original temporal profiles. 
The residual photon waveforms $\xi_{m3}^\mathrm{S,res} (t)$, obtained when applying mode-3 receiver's drive at the optimal timing in the experiment, are shown in Figs.~\ref{fig:5}(b–d).
In this regime, the temporal variation of the incoming photon is slow compared to the receiver response, so the rejection occurs without storing energy in the internal degrees of freedom. Consequently, the scattering is effectively instantaneous, and the reflected photon closely preserves the temporal shape of the input. Therefore, by assigning a receiver mode index larger than that of a sender mode, waveform distortion during rejection can be suppressed.
The squared overlaps among these three rejected modes satisfy $|I_{mm'}|^2\leq0.01$ as summarized in Fig.~\ref{fig:5}(e), indicating that the orthogonality is well preserved.

Furthermore, not only the energy but also the phase information is preserved throughout the mode-selective absorption process. When the modes are matched~($m=n$), the quantum state is transferred from the sender qubit to the receiver qubit. This is verified by reconstructing the experimental process matrix $\chi_\text{exp}$ using quantum process tomography for the two-qubit composite system, yielding an averaged fidelity of $\mathcal{F}=0.62$ across the four modes. The fidelity is primarily limited by photon loss during propagation and finite coherence time but it still exceeds the classical threshold of 0.5 (see Appendix~\ref{sec:app5} for details). When the modes are not matched~($m\neq n$), the quantum state is instead preserved in the rejected photons. A numerical analysis of this process is presented in Appendix~\ref{sec:app4}.

The number of orthogonal modes that can be supported, however, is fundamentally constrained, since the available resources in quantum communication channels, such as the temporal window, frequency bandwidth, and path overhead, are finite in practice. 
Temporal-mode multiplexing is particularly advantageous when both temporal and spectral resources are constrained, as multiple orthogonal modes can be constructed within the same time–frequency space rather than occupying disjoint temporal or spectral regions.
A detailed analysis and comparison with other multiplexing schemes are provided in Appendix~\ref{sec:app3}.
In addition to this, practical limitations on the number of temporal modes may arise from the performance of waveform generators, such as their sampling rate and voltage resolution, which affect the accuracy of the drive pulses.
In practice, these effects are estimated to be about two orders of magnitude smaller than the errors currently observed in the experiment based on our detailed analysis in Appendix~\ref{sec:app2}.

A potential extension of this work is the optimization of the temporal modes.
In the present work, the orthogonal basis was constructed using the hyperbolic secant function.
Designing basis functions that are more resilient to time- and bandwidth-limited photon emission and absorption processes can increase the number of available modes further.
For instance, optimizing the basis through eigenvalue analysis of the transmission spectrum~\cite{gandotra_2025} may offer better localization in both time and frequency domains.

\begin{acknowledgments}
We thank A.~Hern\'andez-Ant\'on, J.-C.~Besse, A.~Wallraff, and A.~Kulikov for sharing their results prior to publication.
We also thank S.~Tamate and H.~Mukai for their fruitful discussions.
This work was partly supported by the Japan Science and Technology Agency (JST) CREST (Grant Number JPMJCR23I4), the JSPS Grant-in-Aid for Scientific Research (KAKENHI) (Grant Number JP22H04937), and the Japan Science and Technology Agency (JST) as part of Adopting Sustainable Partnerships for Innovative Research Ecosystem (ASPIRE) (Grant Number JPMJAP2513).
T.M. acknowledges partial support from the University of Tokyo Forefront Physics and Mathematics Program to Drive Transformation (FoPM), a World-leading Innovative Graduate Study (WINGS) Program.
\end{acknowledgments}

\appendix
\section{Sech-Based Waveform Engineering}
\label{sec:app2}
Following the approach outlined in Ref.~\citenum{penas_multiplexed_2024}, we analytically obtain the orthogonal temporal modes constructed from a hyperbolic secant function.
First, we prepare a set of normalized, linearly independent functions $v_m(t)$~($m=0, 1, 2, \ldots$) as follows:
\begin{equation}
    v_m(t) = \sqrt{N_m} \,\sech(\frac{\kappa_\text{ph} t}{2})\,t^m,
    \label{eq:kappa_ph_def}
\end{equation}
where $\kappa_\text{ph}$ is the pulse-width parameter, and $N_m$ is the normalization constant given by
\begin{equation}
    N_m = \frac{\kappa_\text{ph}^{2m+1}}{8(1-2^{1-2m})\Gamma(2m+1)\zeta(2m)}.
\end{equation}
Here, $\Gamma(k) = (k-1)!$ is the Gamma function for positive integers $k$, and the Riemann zeta function $\zeta(l)$ for even positive integers $l$ is related to the Bernoulli numbers $B_l$ via $\zeta(l)=(-1)^{\frac{l}{2}+1}\frac{(2\pi)^lB_l}{2(l!)}$ . 

\begin{figure}[H]
    \centering
    \includegraphics{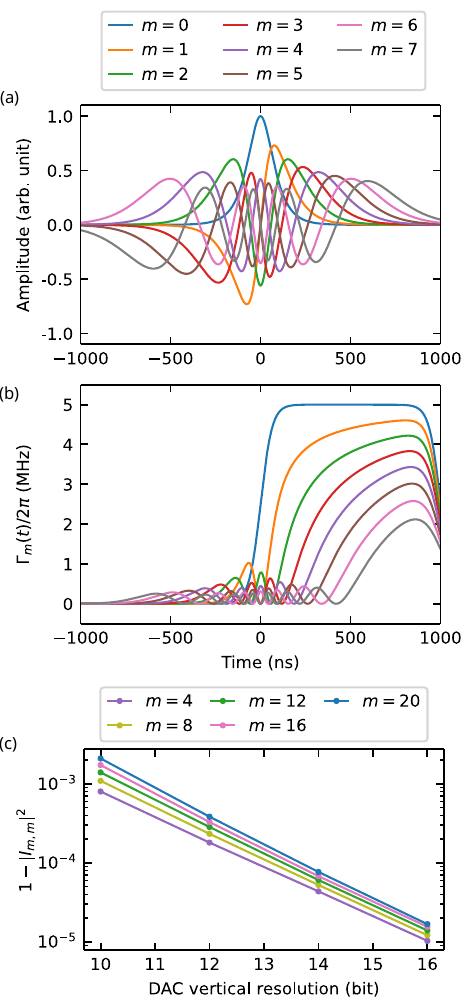}
    \caption{\justifying Sech-based orthogonal waveform engineering. Here, the pulse-width parameter $\kappa_\text{ph}/2\pi$ is set to $5$ MHz.
    (a) Analytically constructed orthogonal photon waveforms $\xi_m(t)$ based on the hyperbolic secant function for $m=0, \ldots,7$.
    (b)~Numerically obtained decay rates $\Gamma_m(t)$ corresponding to $m = 0,\ldots,7$.
    (c) Infidelity of photon waveforms depending on the vertical resolution of the DAC.}
    \label{fig:app2-a}
\end{figure}

To obtain an orthogonal basis in the same subspace, we apply the Gram–Schmidt orthogonalization process:
\begin{equation}
    u_m(t) = v_m(t) - \sum_{k=0}^{m-1} \left(\int_{-\infty}^{\infty}u_k^\ast(t) v_m(t)\,dt\right)u_k(t).
\end{equation}
By normalizing $u_m(t)$, we obtain a set of orthonormalized functions $\{\xi_m(t)\}=\{ u_m(t)/||u_m(t)||\}$.
Due to the inductive nature of this process, each $\xi_m(t)$ can be written analytically as
\begin{equation}
    \xi_m(t) = \sqrt{Z_m} \, \sech(\frac{\kappa_\text{ph} t}{2})A_m(t),
\end{equation}
where $Z_m$ is a normalization constant $Z_m=\sum_{k=0}^m\left(A_{m,m-2k}\sum_{l=0}^m(A_{m,m-2l}/N_{m-(k+l)})\right)$ and $A_m(t)$ is a $m$th-order polynomial defined recursively as
\begin{align}
A_m(t) =\; & t^m - \sum_{i=1}^{\lfloor \frac{m}{2} \rfloor} \left( 
    \int \sech\!\left( \frac{\kappa_\text{ph} t}{2} \right) 
    t^m A_{m-2i}(t)\, dt \right) \notag \\
    & \times Z_{m-2i} A_{m-2i}(t).
\end{align}
When $A_m(t)$ is written in the form $A_m(t) = \sum_{k=0}^m A_{m,k}\,t^k$, we can express it as
\begin{align}
A_m(t) =\; & t^m 
- \sum_{i=1}^{\lfloor \frac{m}{2} \rfloor} \bigg( 
Z_{m-2i} A_{m-2i}(t) \notag \\
& \quad \times \sum_{j=1}^{\lfloor \frac{m}{2} - 1 \rfloor} 
\frac{A_{m-2i,\, m-2(i+j)}}{N_{m-(i+j)}} 
\bigg).
\end{align}
This analytical solution for the orthonormal function set $\{\xi_m(t)\}$, constructed from a hyperbolic secant function, is plotted in Fig.~\ref{fig:app2-a}(a) for the first eight modes ($m=0, \ldots,7$).
We note that the resulting modes are real-valued. The construction can be extended to complex-valued modes by considering the complex span of ${v_m(t)}$.
As the mode order increases, the waveform becomes more complex. 
This corresponds to a broader frequency spectrum, making higher-order modes more susceptible to noise and difficult to be shaped accurately.

Following Eq.~\eqref{eq:2B_Gamma_f}, we can calculate the decay rate $\Gamma_m(t)$ for each photon mode.
Figure~\ref{fig:app2-a}(b) shows the numerically calculated result of $\{\Gamma_m(t)\}$ for $m=0,\ldots,7$.
As the mode order increases, the peak value of $\Gamma(t)$ tends to decrease, which remains achievable from the perspective of device capabilities.
Nevertheless, accurate control may be constrained by the sampling rate and amplitude resolution of the control hardware.
The digital-to-analog converter (DAC) used in our experiment has a 1-GSa/s sampling rate and $12$-bit vertical resolution~($2^{12}$ steps).
While the spectral width of the drive pulses remains below a few tens of megahertz with the current pulse-width parameter $\kappa_\text{ph}$, the limited voltage step size becomes more critical for higher-order modes due to their reduced amplitude.
By quantizing the drive pulses according to the vertical resolution of the DAC, we compare the ideal and discretized waveforms by taking their overlap $I_{mm'}$.
Since the analog bandwidth of the DAC is 400~MHz, the associated smoothing effect is negligible for our pulses, and only the amplitude quantization error is accounted for.
As shown in Fig.~\ref{fig:app2-a}(c), the discretized waveforms approach the ideal ones with increasing bit number, with higher modes being more strongly affected by low amplitude resolution due to the maximum value of the pulse amplitude decreasing with mode number.
Compared with the accuracy of the calibration and bandwidth limitations of the quantum channel, this discretization error is relatively minor, although it may constrain the achievable number of modes in the future.

\section{Experimental Setup}
\label{sec:app1}
\begin{figure*}
    \centering
    \includegraphics{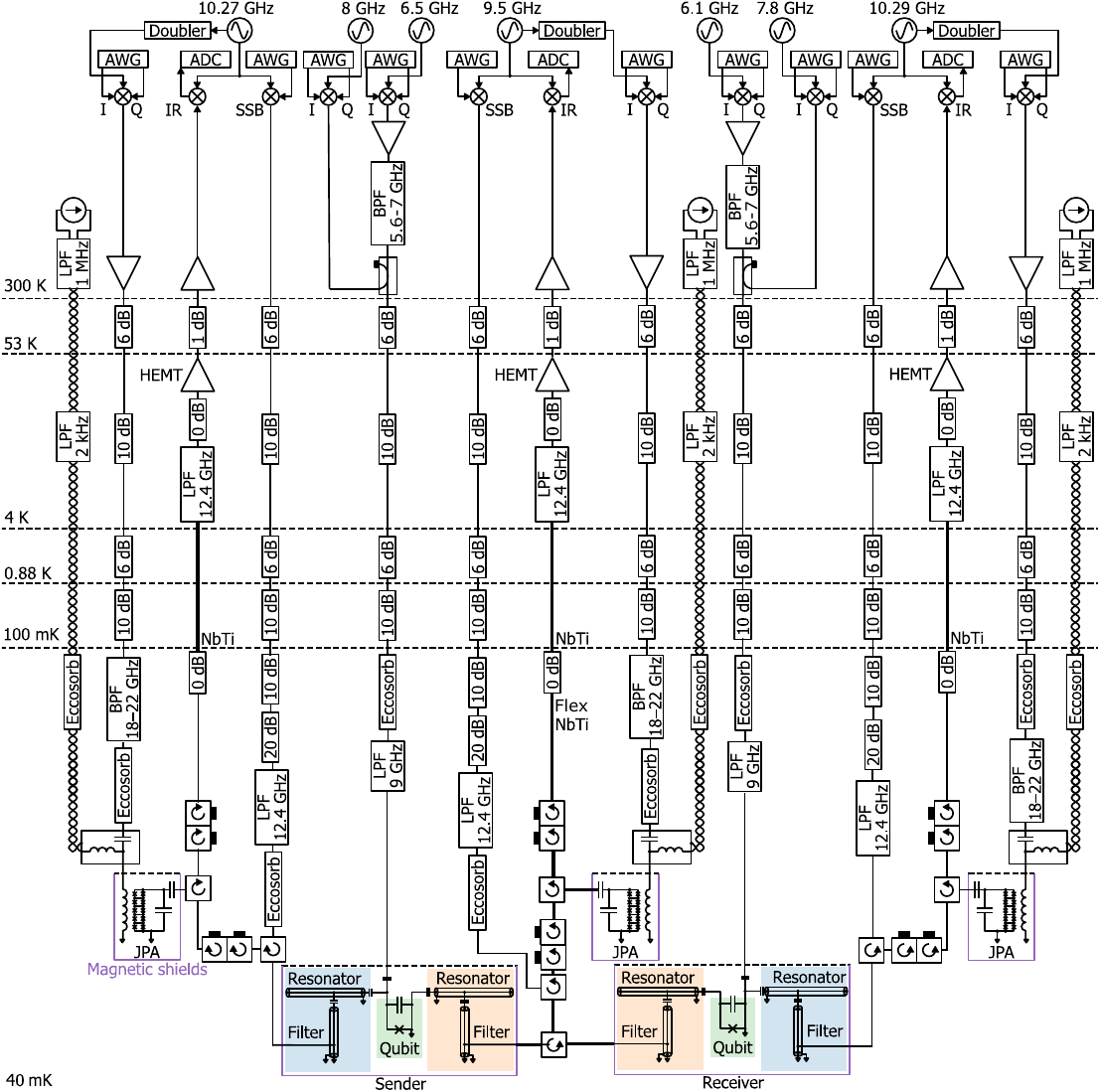}
    \caption{\justifying Cryogenic and room-temperature wiring used in the experiment. The orange-colored resonator--filter systems are used for the photon transfer, while the blue-colored ones are used for qubit readout. AWG, arbitrary waveform generator; ADC, analog-to-digital converter; SSB, single-sideband mixer; IR, image-reject mixer; LPF, low-pass filter; HEMT, high-electron-mobility-transistor amplifier; BPF, band-pass filter; Eccosorb, Eccosorb filter for millimeter-wave-to-infrared radiation; and JPA, flux-driven Josephson parametric amplifier.}
    \label{fig_app1-a}
\end{figure*}

\begin{table*}
  \centering
  \caption{System parameters and coherence of the sender and receiver qubits.}
  \label{tab:qubit_parameters}
  \begin{tabular}{ccc}
    \toprule
    Parameter & Sender & Receiver \\
    \midrule
    Transfer-resonator frequency, $\omega_\mathrm{r}^\text{T}/2\pi$ (GHz) & 9.346 & 9.392 \\
    Transfer-filter frequency, $\omega^\text{T}_\text{f}/2\pi$ (GHz) & 9.339 & 9.394 \\
    Transfer-filter linewidth, $\kappa^\text{T}_\text{f}/2\pi$ (MHz) & 138 & 164 \\
    Transfer-resonator--filter coupling strength, $J^\text{T}/2\pi$ (MHz) & 59 & 50 \\
    Readout-resonator frequency, $\omega_\mathrm{r}^\text{R}/2\pi$ (GHz) & 10.149 & 10.389 \\
    Readout-filter frequency, $\omega^\text{R}_\text{f}/2\pi$ (GHz) & 10.138 & 10.269 \\
    Readout-filter linewidth, $\kappa^\text{R}_\text{f}/2\pi$ (MHz) & 35 & 34 \\
    Readout-resonator--filter coupling strength, $J^\text{R}/2\pi$ (MHz) & 43 & 64 \\
    Qubit frequency, $\omega_{ge}/2\pi$ (GHz) & 8.207 & 7.977 \\
    Anharmonicity, $\alpha/2\pi$ (MHz) & $-356$ & $-356$ \\
    \midrule
    Energy-relaxation time of $|e\rangle$, $T_{1ge}$ ($\mu$s) & 29 & 19 \\
    Energy-relaxation time of $|f\rangle$, $T_{1ef}$ ($\mu$s) & 22 & 11 \\
    Dephasing time of $|e\rangle$, $T_{2ge}^*$ ($\mu$s) & 11 & 9 \\
    \bottomrule
  \end{tabular}
  \label{tb:app1-b}
\end{table*}

The devices used in this work consist of a fixed-frequency transmon qubit and two resonator--filter systems.
One system is dedicated to photon transfer, while the other is used for qubit readout.
A resonator integrated with an intrinsic Purcell filter~\cite{Sunada_intrinsic_purcell_2022, Spring_intrinsic_purcell_2025} and an additional band-pass filter~\cite{Sete_bandpass_purcell_2015, Jeffrey_bandpass_purcell_2014} enables fast photon emission and absorption and broad photon bandwidth while suppressing qubit energy relaxation.
The relevant device parameters and measured coherence times are listed in Table~\ref{tb:app1-b}. Further details on the device design can be found in Ref.~\cite{TMiyamura_qcomm_2025}.

A schematic of the setup is shown in Fig.~\ref{fig_app1-a}.
Each device is mounted in a separate magnetic shield within a single dilution refrigerator.
Their transfer ports are connected via a flexible NbTi coaxial cable (Koaxis, CC086NBTI).
The reflection from the receiver is routed via a circulator to a JPA, where the photon waveform is amplified.
The flexible cable employs a niobium-titanium-alloy center conductor with a silver-plated copper-braid shield and PTFE dielectric, which is expected to minimize photon loss in the cables.
Inside the magnetic shields, however, the cable sections cannot be replaced, and approximately 15-cm-long copper semi-flexible cables are used in those regions.
Additional losses may occur at the gold-plated SMA connectors used to join the cables and connect the devices.
The readout ports are connected to a JPA, enabling single-shot readout of the qubit state.

\section{Process Tomography of the Quantum State Transfer}
\label{sec:app5}
\begin{figure}[H]
    \centering
    \includegraphics{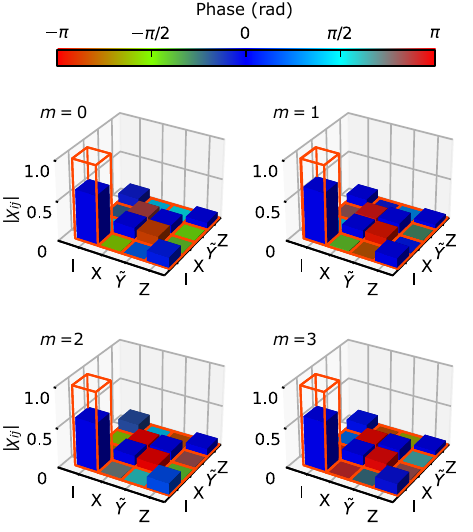}
    \caption{\justifying Reconstructed process matrices in the basis $\{I, \,X, \,\tilde{Y}, \,Z\}$ for quantum state transfer using temporal modes $m=0,\,\ldots,\,3$. 
    Colored bars represent the experimental result $\chi_\text{exp}$, while coral outlines indicate the target matrices $\chi_\text{ideal}$.}
    \label{fig:app4_a}
\end{figure}

We evaluate the quantum state transfer for the cases with $m=n$ with quantum process tomography by reconstructing the process matrix $\chi_\text{exp}$.
To obtain $\chi_\text{exp}$, we prepare the six mutually unbiased qubit basis states at the sender, $\ket{\psi}^\text{S}$: $\ket{g}$, $\ket{e}$, $(\ket{g}+\ket{e})/\sqrt{2}$, $(\ket{g}+i\ket{e})/\sqrt{2}$, $(\ket{g}-\ket{e})/\sqrt{2}$, and $(\ket{g}-i\ket{e})/\sqrt{2}$.
These states are transferred to the receiver using itinerant microwave photons in temporal-mode $m=0,\,1,\,2,$ and $3$.
At the receiver, we perform quantum state tomography with single-shot readout.
The receiver is treated as a qutrit system, and its final state, $\ket{\psi}^\text{R}$ is measured in different bases using the gate operations $\textbf{I}$, $(\pi/2)^x_{ge}$, $(\pi/2)^y_{ge}$, $\pi^x_{ge}$, $(\pi/2)^x_{ef}$, $(\pi/2)^y_{ef}$, $\pi^x_{ge}(\pi/2)^x_{ef}$, $\pi^x_{ge}(\pi/2)^y_{ef}$, and $\pi^x_{ge}\pi^x_{ef}$, where $\theta^a_{ij}$ denotes a pulse of angle $\theta$ on the $ij$ transition along the $a$-axis~\cite{RBianchetti_QPT_2010}.
The final states are reconstructed with a maximum-likelihood method, ensuring that the obtained density matrices are valid.
From these reconstructed states, we obtain the process matrices $\chi_\text{exp}$ in the basis of the Pauli matrices $I$, $X=\sigma_x$, $\tilde{Y}=i\sigma_y$, and $Z=\sigma_z$, as shown in Fig.~\ref{fig:app4_a}.
The corresponding process fidelities are $\mathcal{F}_p=\text{Tr}[\chi_\text{exp}\chi_\text{ideal}] = 0.62,\,0.65,\,0.61,$ and $0.61$ for $m=0,\,1,\,2,$ and $3$, respectively.

The main factor limiting the fidelity is photon loss during propagation ($\sim$17\%), including losses in the coaxial cables, circulator, and connectors.
The photon loss is estimated from the emitted and received photon fluxes as $l=1-\int|\xi_m^\text{S}(t)|^2dt/\int|\xi_n^\text{R}(t)|^2dt$.
The average photon loss across the four modes is $0.33$, which could be overestimated due to the unexpected leakage during photon emission~\cite{TMiyamura_qcomm_2025}.
Other error sources include the finite coherence times of the qubits ($\sim$7\%), imperfect absorption efficiency (1--5\%), the readout infidelity ($\sim$3\%), and possible calibration errors of $\pi_\text{ef}$ pulses.

\section{Master-Equation Modeling of Mode-Selective Quantum State Transfer}
\label{sec:app4}
\subsection{Quantum-system–traveling-field interaction}
\begin{figure}
    \centering
    \includegraphics{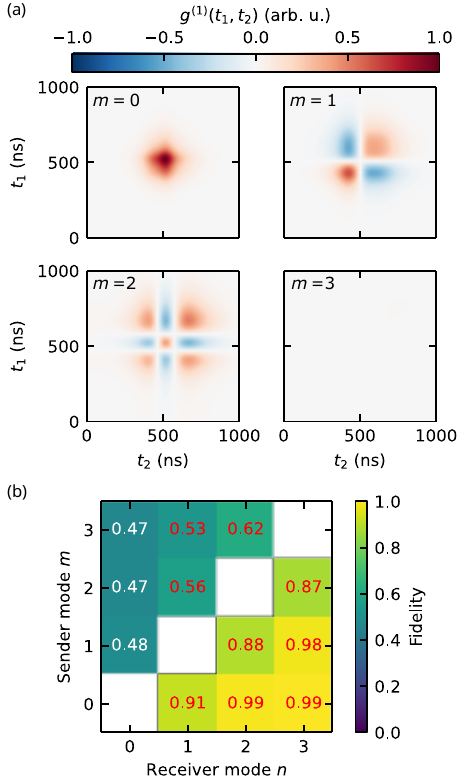}
    \caption{\justifying 
    (a)~Autocorrelation of the output field when implementing mode-selective absorption for photon modes $m=0,\,\ldots,\,3$ with the receiver mode $n=3$.
    (b)~Simulated fidelities of the quantum state transferred from the incoming photon to the rejected photon for sender and receiver modes $m,n=0,\,\ldots,\,3$ with $m\neq n$.
    The rejected photons are absorbed by the second virtual cavity designed to absorb the incoming-photon waveform $\xi_m(t)$.
    }
    \label{fig:app5_a}
\end{figure}

We model the receiver as a coupled system consisting of an anharmonic oscillator (transmon, operator $\hat{a}$) and a harmonic oscillator (resonator, operator $\hat{b}$).  
We adopt units where $\hbar = 1$.
The effective Hamiltonian of this system under the receiver's drive~$\zeta^{\text{R}}_n(-t+\Delta t)$ is~\cite{kurpiers_deterministic_2018}
\begin{align}
    \hat{\mathcal{H}}_\text{eff}&=-\frac{\alpha}{2}\hat{b}^\dagger\hat{b}+\frac{\alpha}{2}\hat{b}^\dagger\hat{b}^\dagger\hat{b}\hat{b}+\frac{K}{2}\hat{a}^\dagger\hat{a}^\dagger\hat{a}\hat{a}+2\chi\hat{a}^\dagger\hat{a}\hat{b}^\dagger\hat{b}\notag \\
    &+\Bigl(g_n^{f0g1}(-t+\Delta t)\hat{b}^\dagger\hat{b}^\dagger\hat{a}+g_n^{f0g1\ast}(-t+\Delta t)\hat{b}\hat{b}\hat{a}^\dagger\Bigr),
\end{align}
where $g_n^{f0g1}$ represents the coupling  between $\ket{f,0}$ and $\ket{g,1}$:
\begin{equation}
    g_n^{f0g1}(-t+\Delta t)=\frac{\sqrt{\kappa_\mathrm{f}^\text{T}}}{2}\frac{\xi_n( t)}{\sqrt{\int_0^{t} |\xi_n(\tau)|^2d\tau}}.
\end{equation}
Here, $\alpha$ denotes the qubit anharmonicity, $K$ the resonator anharmonicity induced by the qubit, and $\chi$ the dispersive shift.
The traveling field incident on the receiver is modeled as the output field of a virtual cavity with operator $\hat{c}_\xi$~\cite{AHKiilerich_vc_2020}. 
The coupling term of the Hamiltonian between the receiver and the virtual cavity is
\begin{equation}
    \hat{\mathcal{H}}_\mathrm{ph}=i(g_m^\mathrm{ph}(t)\hat{c}_\xi^\dagger\hat{a}-g_m^{\mathrm{ph}\ast}(t)\hat{c}_\xi\hat{a}^\dagger),
\end{equation}
where
\begin{equation}
    g_m^\mathrm{ph}(t)=\frac{\sqrt{\kappa_\mathrm{f}^\text{T}}}{2}\frac{\xi_m^\ast(t)}{\sqrt{1-\int_0^t |\xi_m(\tau)|^2d\tau}}.
\end{equation}
The system dynamics are governed by the master equation
\begin{equation}
    \dot\rho = -i[\hat{\mathcal{H}}_\text{eff}+\hat{\mathcal{H}}_\mathrm{ph}, \rho]+\sum_{i=0}^n\mathcal{D}[\hat{L}_i]\rho,
    \label{eq:master_eq}
\end{equation}
where the Lindblad operator
\begin{equation}
    \hat{L}_0=\frac{2}{\sqrt{\kappa_\mathrm{f}^\text{T}}}g_m^{\mathrm{ph}\ast}(t)\hat{c}_\xi+\sqrt{\kappa_\mathrm{f}^\text{T}}\, \hat{a}
\end{equation}
describes the interference between the incoming itinerant photon and the outgoing field of the receiver, and the additional $\hat{L}_i$ account for qubit decay and dephasing.
Here, the dissipator is defined as
\begin{equation}
\mathcal{D}[\hat{L}] \rho
= \hat{L} \rho \hat{L}^\dagger
- \frac{1}{2}\left( \hat{L}^\dagger \hat{L}\rho + \rho \hat{L}^\dagger \hat{L} \right).
\end{equation}

Finally, the time-dependent intensity of the output field is given by
\begin{equation}
    I_\text{out}(t)=\langle[\hat{L}_0(t)]^\dagger\hat{L}_0(t)\rangle.
\end{equation}
We can also compute the autocorrelation of the output field.
Note that the initial state of the virtual cavity is prepared as the superposition of the ground and excited states, corresponding to the Fock-state of the traveling field in the experiment.
The corresponding fields, obtained when implementing mode-selective absorption for photons in modes $m=0, \ldots,3$ with the receiver mode $n=3$, are shown in Fig.~\ref{fig:app5_a}(a). Its eigenmode decomposition,
\begin{equation}
    g^{(1)}(t_1, t_2)=\langle[\hat{L}_0(t_1)]^\dagger\hat{L}_0(t_2)\rangle=\sum_in_iv_i^\ast(t_1) v_i(t_2)
\end{equation}
yields a set of orthogonal modes $v_i(t)$ with corresponding occupancies $n_i$.
The dominant mode $v_0$ has a high occupancy for $m=0, \,1,$ and~2~($\ne n=3$), with $n_0=1.000,\,0.999,$ and $0.995$, respectively. 
This indicates that the rejected photons largely retain their single-photon character, allowing mode-selective absorption at a downstream receiver.
The measured residual photon fluxes presented in the main text are well captured by the dominant mode $v_0$, as shown in Figs.~\ref{fig:5}(b--d).

\subsection{Preservation of photon waveforms and quantum states in mode-selective absorption}
\begin{figure}
    \centering
    \includegraphics{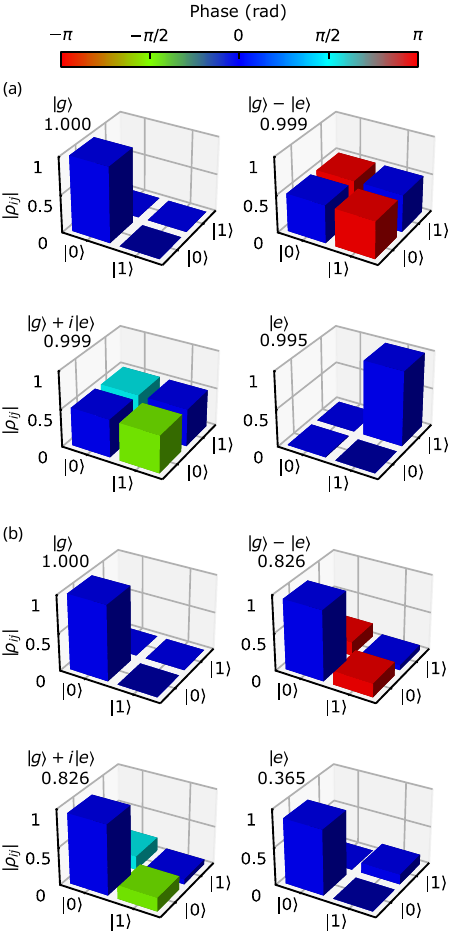}
    \caption{\justifying 
    Density matrices $\rho$ of the second virtual cavity after absorbing the rejected photon for mode combinations (a)~$(m,n)=(0,3)$ and (b)~$(m,n)=(3,0)$.
    For each case, the results are shown for four different initial states of the first virtual cavity: $|g\rangle$, $(|g\rangle-|e\rangle)/\sqrt{2}$, $(|g\rangle+i|e\rangle)/\sqrt{2}$, and $|e\rangle$. The value shown below each state label indicates the corresponding fidelity.
    }
    \label{fig:app5_b}
\end{figure}

The quantum state transfer for the absorbed mode~($m=n$) can be evaluated from the final state of the receiver qubit, as performed in the experiment via quantum process tomography.
To assess the preservation of the quantum state against rejected modes ($m\neq n$), we introduce a second virtual cavity in the simulation, placed downstream of the receiver, with operator $\hat{c}_v$~\cite{AHKiilerich_vc_2020}.
This cavity is configured to absorb the rejected photons by coupling it to the receiver with the time-reversed waveform $g_m^{\mathrm{ph}\ast}(-t+\Delta t)$, the one that would capture the original photon waveform $\xi(t)$.
The additional Hamiltonian to Eq.~\eqref{eq:master_eq} is
\begin{equation}
    \hat{\mathcal{H}}_v=i\left(g_m^{\mathrm{ph}\ast}(-t+\Delta t)\hat{c}_v^\dagger\hat{a}-g_m^{\mathrm{ph}}(-t+\Delta t)\hat{c}_v\hat{a}^\dagger\right),
\end{equation}
and the Lindblad operator is 
\begin{equation}
    \hat{L}_0=\frac{2}{\sqrt{\kappa_\mathrm{f}^\text{T}}}g_m^{\mathrm{ph}\ast}(t)\hat{c}_\xi+\frac{2}{\sqrt{\kappa_\mathrm{f}^\text{T}}}g_m^{\mathrm{ph}}(-t+\Delta t)\hat{c}_v+\sqrt{\kappa_\mathrm{f}^\text{T}}\, \hat{a}.
\end{equation}
By evaluating the quantum state stored in the second cavity, we quantify both the preservation of the photon waveform and the quantum state.

We prepare the initial state in the first virtual cavity and emit a photon in temporal mode $m$.
When the photon interacts with the receiver which is driven to absorb mode $n$, it is rejected for $m\neq n$ and subsequently captured by the second virtual cavity designed to absorb a mode-$m$ photon.
The simulated fidelities of the quantum state transfer from the incoming photon to the rejected photon show higher values for $m<n$, whereas they are significantly decreased for $m>n$, as displayed in Fig.~\ref{fig:app5_a}(b).
The final states of the second cavity are shown in Figs.~\ref{fig:app5_b}(a) and~(b) for $(m,n)=(0,3)$ and $(3,0)$, which correspond to the highest and lowest fidelities, respectively.
In these simulations, we take into account the finite coherence time of the receiver qubit.
Even in the low-fidelity case ($m>n$), the phase information in the first cavity is still transferred to the second cavity.
This preservation of phase information indicates that the reduced fidelity arises not from the loss of quantum coherence but from waveform distortion, which prevents efficient absorption by the second cavity.
By choosing modes such that $m<n$ or by redesigning the receiver coupling strength using $v_0$, selective quantum state transfer across multiple receivers can be achieved.

\section{Constraints on the Number of Orthogonal Modes Under Finite Time- and Bandwidth-Resources}
\label{sec:app3}
\subsection{Time--frequency distribution of orthogonal pulses}
\begin{figure}
    \centering
    \includegraphics{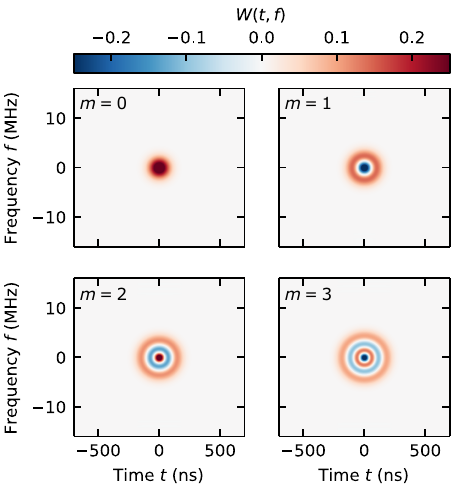}
    \caption{\justifying Chronocyclic Wigner distributions of orthogonal Hermite-Gaussian pulses for modes $m=0,\,\ldots,\,3$.
    }
    \label{fig:app3_f}
\end{figure}

The distribution of pulses in the time–frequency space is characterized by the chronocyclic Wigner function~\cite{Paye_chronocyclic_Wigner_1992}:
\begin{equation}
    W(t, f)= \int \!d\tau\,\xi\left(t+\frac{\tau}{2}\right)\xi^\ast\left(t-\frac{\tau}{2}\right)\exp(-i2\pi f\tau).
\end{equation}
The distribution captures the temporal and spectral characteristics of the mode functions.
For example, the time–frequency distribution of Hermite–Gaussian modes exhibits a concentric structure~(Fig.~\ref{fig:app3_f}),
whereas the diamond-like shape of the orthogonal modes constructed from the hyperbolic secant function [Fig.~\ref{fig:4}(a)] originates from their long temporal tails.

The overlap of the distributions of orthogonal modes vanishes:
\begin{equation}
    \int \!df\!\int \!dt\, W_m(t, f)W_{m'}(t, f)=\delta_{mm'}
    \label{eq:wigner_overlap}
\end{equation}
In temporal-mode multiplexing, although the orthogonal modes occupy the same time–frequency space, Eq.~\eqref{eq:wigner_overlap} is still satisfied because negative regions in the Wigner function ensure their orthogonality.

\subsection{Number of available orthogonal modes}
\begin{figure*}
    \centering
    \includegraphics{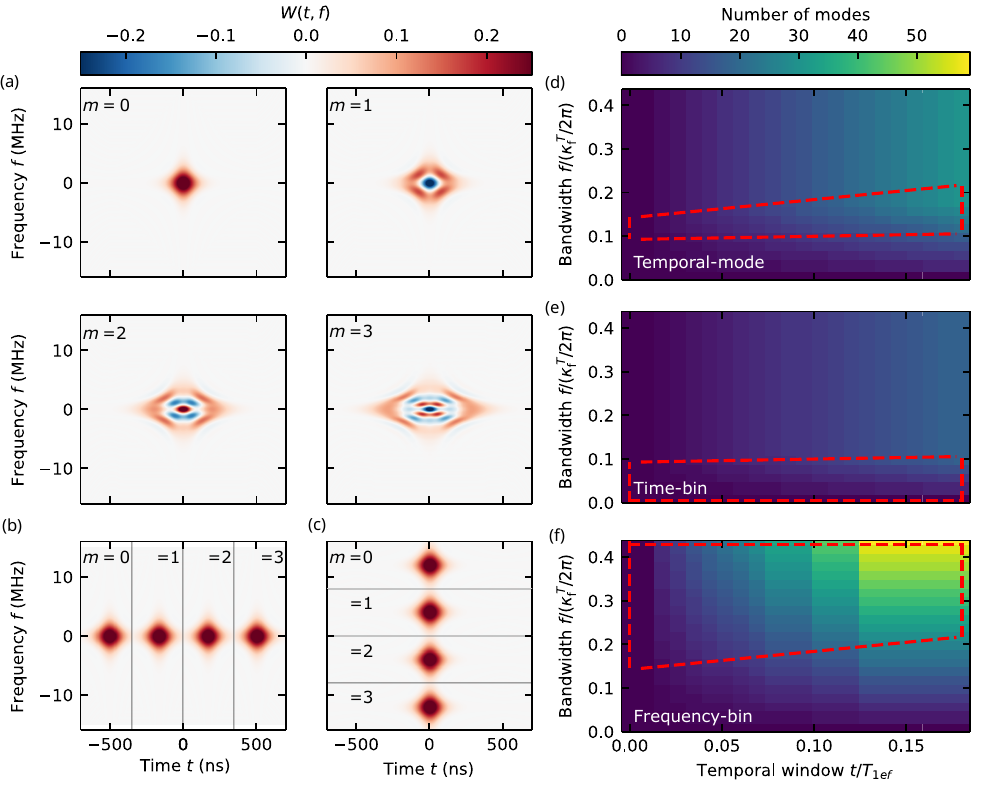}
    \caption{\justifying 
    Comparison of orthogonal modes used in temporal-mode, time-bin, and frequency-bin multiplexing schemes under finite temporal and spectral constraints. 
    (a--c)~Time--frequency Wigner distributions of orthogonal pulses for modes $m = 0,\,\ldots,\,3$ for each multiplexing scheme.
    (d--f)~Number of modes supported within the given temporal window and bandwidth.
    The horizontal axis is normalized by the sender energy relaxation time of $|f\rangle$, $T_{1ef}$, while the vertical axis is normalized by the linewidth of the sender transfer filter resonator, $\kappa_\mathrm{f}^\mathrm{T}/2\pi$.
    Red dashed lines indicate the advantageous region of each multiplexing scheme:
    (a),(d)~temporal-mode,
    (b),(e)~time-bin, and
    (c),(f)~frequency-bin multiplexing.
    }
    \label{fig:4}
\end{figure*}

We examine how the finite time and frequency resources in the quantum channel affect the number of practically available temporal modes and the advantages of temporal-mode multiplexing compared with alternative approaches, such as time- and frequency-bin multiplexing.
We define the temporal window and bandwidth required for each mode as the regions containing 99\% of its total energy in the respective domains.
Analytically derived temporal modes are used for temporal-mode multiplexing (Appendix~\ref{sec:app2}), whereas time- or frequency-bin modes are constructed by sequentially arranging the waveform of the mode\nobreakdash-0 temporal-mode in the time or frequency domain, respectively.
Their corresponding distributions in a time--frequency space are shown in Figs.~\ref{fig:4}(a--c) using chronocyclic Wigner functions.

The pulse-width parameter $\kappa_\text{ph}$ determines the temporal and spectral widths of the pulses in this space, with increasing $\kappa_\text{ph}$ narrowing the pulse in time while broadening it in frequency, and vice versa. Its optimal value depends on the multiplexing scheme.
The upper bound of $\kappa_\text{ph}$ is limited by the device’s decay rate (set to $8$ MHz here), whereas the lower bound is constrained by the temporal window available in the experiment. This temporal window is limited by the coherence of the sender and receiver due to their internal loss, which reduces the achievable fidelity of photon generation.
For temporal-mode multiplexing, the number of accessible modes increases with both the temporal and spectral resources~[Fig.~\ref{fig:4}(d)].
For time-bin and frequency-bin multiplexing, it grows primarily with the temporal window and frequency bandwidth, respectively~[Figs.~4(e) and~(f)].
Note that frequency-bin multiplexing also depends on the temporal window, since a longer window allows a smaller $\kappa_\text{ph}$, thus narrowing the spectrum.

Each multiplexing scheme shows an advantage in different regions of the time–frequency space, as shown by the red dashed lines in Figs.~\ref{fig:4}(d--f). While time-bin and frequency-bin multiplexing become advantageous when a longer temporal window or a wider bandwidth is available, temporal-mode multiplexing is advantageous in the intermediate region between them.
In future setups involving multiple senders and receivers, where device coherence times and shared communication bandwidth impose stricter constraints in both time and frequency, temporal-mode multiplexing can offer an advantage.

As an extension of this analysis, we can consider combining multiple multiplexing approaches.
For example, the combination of time- and frequency-bin multiplexing increases the number of orthogonal modes in a two-dimensional mode space.
Furthermore, different multiplexing degrees of freedom can be assigned distinct roles.
One can be used to expand the Hilbert-space dimension of quantum network channels, while another can be employed for dual-rail encoding, enabling photon-loss detection.

\FloatBarrier
\nocite{*}
\bibliography{bibliography}

@article{axline_-demand_2018,
title = {On-demand quantum state transfer and entanglement between remote microwave cavity memories},
volume = {14},
issn = {1745-2473, 1745-2481},
url = {http://www.nature.com/articles/s41567-018-0115-y},
doi = {10.1038/s41567-018-0115-y},
number = {7},
urldate = {2022-11-01},
journal = {Nature Phys},
author = {Axline, Christopher J. and Burkhart, Luke D. and Pfaff, Wolfgang and Zhang, Mengzhen and Chou, Kevin and Campagne-Ibarcq, Philippe and Reinhold, Philip and Frunzio, Luigi and Girvin, S. M. and Jiang, Liang and Devoret, M. H. and Schoelkopf, R. J.},
month = jul,
year = {2018},
pages = {705--710},
}

@article{pechal_microwave-controlled_2014,
title = {Microwave-Controlled Generation of Shaped Single Photons in Circuit Quantum Electrodynamics},
volume = {4},
issn = {2160-3308},
url = {https://link.aps.org/doi/10.1103/PhysRevX.4.041010},
doi = {10.1103/PhysRevX.4.041010},
number = {4},
urldate = {2022-10-28},
journal = {Phys. Rev. X},
author = {Pechal, M. and Huthmacher, L. and Eichler, C. and Zeytino\u{g}lu, S. and Abdumalikov, A. A. and Berger, S. and Wallraff, A. and Filipp, S.},
month = oct,
year = {2014},
pages = {041010},
}

@article{zeytinoglu_microwave-induced_2015,
title = {Microwave-induced amplitude- and phase-tunable qubit-resonator coupling in circuit quantum electrodynamics},
volume = {91},
issn = {1050-2947, 1094-1622},
url = {https://link.aps.org/doi/10.1103/PhysRevA.91.043846},
doi = {10.1103/PhysRevA.91.043846},
number = {4},
urldate = {2022-11-01},
journal = {Phys. Rev. A},
author = {Zeytino\u{g}lu, S. and Pechal, M. and Berger, S. and Abdumalikov, A. A. and Wallraff, A. and Filipp, S.},
month = apr,
year = {2015},
pages = {043846},
}

@article{fowler_surface_2012,
title = {Surface codes: {Towards} practical large-scale quantum computation},
volume = {86},
copyright = {http://link.aps.org/licenses/aps-default-license},
issn = {1050-2947, 1094-1622},
shorttitle = {Surface codes},
url = {https://link.aps.org/doi/10.1103/PhysRevA.86.032324},
doi = {10.1103/PhysRevA.86.032324},
number = {3},
urldate = {2024-05-02},
journal = {Phys. Rev. A},
author = {Fowler, Austin G. and Mariantoni, Matteo and Martinis, John M. and Cleland, Andrew N.},
month = sep,
year = {2012},
pages = {032324},
}

@article{grebel_bidirectional_2024,
title = {Bidirectional Multiphoton Communication between Remote Superconducting Nodes},
volume = {132},
issn = {0031-9007, 1079-7114},
url = {https://link.aps.org/doi/10.1103/PhysRevLett.132.047001},
doi = {10.1103/PhysRevLett.132.047001},
number = {4},
urldate = {2024-03-27},
journal = {Phys. Rev. Lett.},
author = {Grebel, Joel and Yan, Haoxiong and Chou, Ming-Han and Andersson, Gustav and Conner, Christopher R. and Joshi, Yash J. and Miller, Jacob M. and Povey, Rhys G. and Qiao, Hong and Wu, Xuntao and Cleland, Andrew N.},
month = jan,
year = {2024},
pages = {047001},
}

@article{storz_loophole-free_2023,
title = {Loophole-free {Bell} inequality violation with superconducting circuits},
volume = {617},
copyright = {2023 The Author(s)},
issn = {1476-4687},
url = {https://www.nature.com/articles/s41586-023-05885-0},
doi = {10.1038/s41586-023-05885-0},
number = {7960},
urldate = {2024-02-07},
journal = {Nature},
author = {Storz, Simon and Schär, Josua and Kulikov, Anatoly and Magnard, Paul and Kurpiers, Philipp and Lütolf, Janis and Walter, Theo and Copetudo, Adrian and Reuer, Kevin and Akin, Abdulkadir and Besse, Jean-Claude and Gabureac, Mihai and Norris, Graham J. and Rosario, Andrés and Martin, Ferran and Martinez, José and Amaya, Waldimar and Mitchell, Morgan W. and Abellan, Carlos and Bancal, Jean-Daniel and Sangouard, Nicolas and Royer, Baptiste and Blais, Alexandre and Wallraff, Andreas},
month = may,
year = {2023},
pages = {265--270},
}

@article{hertzberg_laser-annealing_2021,
title = {Laser-annealing {Josephson} junctions for yielding scaled-up superconducting quantum processors},
volume = {7},
issn = {2056-6387},
url = {https://www.nature.com/articles/s41534-021-00464-5},
doi = {10.1038/s41534-021-00464-5},
number = {1},
urldate = {2023-12-01},
journal = {npj Quantum Inf},
author = {Hertzberg, Jared B. and Zhang, Eric J. and Rosenblatt, Sami and Magesan, Easwar and Smolin, John A. and Yau, Jeng-Bang and Adiga, Vivekananda P. and Sandberg, Martin and Brink, Markus and Chow, Jerry M. and Orcutt, Jason S.},
month = aug,
year = {2021},
pages = {129},
}

@article{koch_charge-insensitive_2007,
title = {Charge-insensitive qubit design derived from the {Cooper} pair box},
volume = {76},
issn = {1050-2947, 1094-1622},
url = {https://link.aps.org/doi/10.1103/PhysRevA.76.042319},
doi = {10.1103/PhysRevA.76.042319},
number = {4},
urldate = {2023-01-29},
journal = {Phys. Rev. A},
author = {Koch, Jens and Yu, Terri M. and Gambetta, Jay and Houck, A. A. and Schuster, D. I. and Majer, J. and Blais, Alexandre and Devoret, M. H. and Girvin, S. M. and Schoelkopf, R. J.},
month = oct,
year = {2007},
pages = {042319},
}

@article{campagne-ibarcq_deterministic_2018,
title = {Deterministic Remote Entanglement  of Superconducting Circuits through Microwave Two-Photon Transitions},
volume = {120},
issn = {0031-9007, 1079-7114},
url = {https://link.aps.org/doi/10.1103/PhysRevLett.120.200501},
doi = {10.1103/PhysRevLett.120.200501},
number = {20},
urldate = {2023-09-10},
journal = {Phys. Rev. Lett.},
author = {Campagne-Ibarcq, P. and Zalys-Geller, E. and Narla, A. and Shankar, S. and Reinhold, P. and Burkhart, L. and Axline, C. and Pfaff, W. and Frunzio, L. and Schoelkopf, R. J. and Devoret, M. H.},
month = may,
year = {2018},
pages = {200501},
}

@article{ang_architectures_2022,
author = {Ang, James and Carini, Gabriella and Chen, Yanzhu and Chuang, Isaac and Demarco, Michael and Economou, Sophia and Eickbusch, Alec and Faraon, Andrei and Fu, Kai-Mei and Girvin, Steven and Hatridge, Michael and Houck, Andrew and Hilaire, Paul and Krsulich, Kevin and Li, Ang and Liu, Chenxu and Liu, Yuan and Martonosi, Margaret and McKay, David and Misewich, Jim and Ritter, Mark and Schoelkopf, Robert and Stein, Samuel and Sussman, Sara and Tang, Hong and Tang, Wei and Tomesh, Teague and Tubman, Norm and Wang, Chen and Wiebe, Nathan and Yao, Yongxin and Yost, Dillon and Zhou, Yiyu},
title = {ARQUIN: Architectures for Multinode Superconducting Quantum Computers},
year = {2024},
issue_date = {September 2024},
publisher = {Association for Computing Machinery},
address = {New York, NY, USA},
journal = {ACM Transactions on Quantum Computing},
volume = {5},
number = {3},
url = {https://doi.org/10.1145/3674151},
doi = {10.1145/3674151},
month = sep,
articleno = {19},
pages = {1--59},
numpages = {59}
}

@article{kurpiers_deterministic_2018,
title = {Deterministic quantum state transfer and remote entanglement using microwave photons},
volume = {558},
issn = {0028-0836, 1476-4687},
url = {http://www.nature.com/articles/s41586-018-0195-y},
doi = {10.1038/s41586-018-0195-y},
number = {7709},
urldate = {2022-11-01},
journal = {Nature},
author = {Kurpiers, P. and Magnard, P. and Walter, T. and Royer, B. and Pechal, M. and Heinsoo, J. and Salathé, Y. and Akin, A. and Storz, S. and Besse, J.-C. and Gasparinetti, S. and Blais, A. and Wallraff, A.},
month = jun,
year = {2018},
pages = {264--267},
}

@article{Leung_deterministic_2019,
title = {Deterministic bidirectional communication and remote entanglement generation between superconducting qubits},
volume = {5},
number = {18},
url = {https://doi.org/10.1038/s41534-019-0128-0},
doi = {10.1038/s41534-019-0128-0},
journal = {npj Quantum Inf},
author = {N. Leung and Y. Lu and S. Chakram and R. K. Naik and N. Earnest and R. Ma and K. Jacobs and A. N. Cleland and D. I. Schuster },
month = feb,
year = {2019},
}

@article{ilves_-demand_2020,
title = {On-demand generation and characterization of a microwave time-bin qubit},
volume = {6},
issn = {2056-6387},
url = {http://www.nature.com/articles/s41534-020-0266-4},
doi = {10.1038/s41534-020-0266-4},
number = {1},
urldate = {2022-11-02},
journal = {npj Quantum Inf},
author = {Ilves, J. and Kono, S. and Sunada, Y. and Yamazaki, S. and Kim, M. and Koshino, K. and Nakamura, Y.},
month = dec,
year = {2020},
pages = {34},
}

@article{kurpiers_quantum_2019,
title = {Quantum Communication with Time-Bin Encoded Microwave Photons},
volume = {12},
issn = {2331-7019},
url = {https://link.aps.org/doi/10.1103/PhysRevApplied.12.044067},
doi = {10.1103/PhysRevApplied.12.044067},
number = {4},
urldate = {2022-11-21},
journal = {Phys. Rev. Applied},
author = {Kurpiers, P. and Pechal, M. and Royer, B. and Magnard, P. and Walter, T. and Heinsoo, J. and Salathé, Y. and Akin, A. and Storz, S. and Besse, J.-C. and Gasparinetti, S. and Blais, A. and Wallraff, A.},
month = oct,
year = {2019},
pages = {044067},
}

@article{kannan_generating_2020,
title = {Generating spatially entangled itinerant photons with waveguide quantum electrodynamics},
volume = {6},
copyright = {https://creativecommons.org/licenses/by-nc/4.0/},
issn = {2375-2548},
url = {https://www.science.org/doi/10.1126/sciadv.abb8780},
doi = {10.1126/sciadv.abb8780},
number = {41},
urldate = {2024-09-16},
journal = {Sci. Adv.},
author = {Kannan, B. and Campbell, D. L. and Vasconcelos, F. and Winik, R. and Kim, D. K. and Kjaergaard, M. and Krantz, P. and Melville, A. and Niedzielski, B. M. and Yoder, J. L. and Orlando, T. P. and Gustavsson, S. and Oliver, W. D.},
month = oct,
year = {2020},
pages = {eabb8780},
}

@article{penas_multiplexed_2024,
title = {Multiplexed quantum state transfer in waveguides},
volume = {6},
issn = {2643-1564},
url = {https://link.aps.org/doi/10.1103/PhysRevResearch.6.033294},
doi = {10.1103/PhysRevResearch.6.033294},
number = {3},
urldate = {2024-09-21},
journal = {Phys. Rev. Research},
author = {Peñas, Guillermo F. and Puebla, Ricardo and García-Ripoll, Juan José},
month = sep,
year = {2024},
pages = {033294},
}

@article{kannan_waveguide_2020,
title = {Waveguide quantum electrodynamics with superconducting artificial giant atoms},
volume = {583},
copyright = {2020 The Author(s), under exclusive licence to Springer Nature Limited},
issn = {1476-4687},
url = {https://www.nature.com/articles/s41586-020-2529-9},
doi = {10.1038/s41586-020-2529-9},
number = {7818},
urldate = {2024-09-21},
journal = {Nature},
author = {Kannan, Bharath and Ruckriegel, Max J. and Campbell, Daniel L. and Frisk Kockum, Anton and Braum\"uller, Jochen and Kim, David K. and Kjaergaard, Morten and Krantz, Philip and Melville, Alexander and Niedzielski, Bethany M. and Veps\"al\"ainen, Antti and Winik, Roni and Yoder, Jonilyn L. and Nori, Franco and Orlando, Terry P. and Gustavsson, Simon and Oliver, William D.},
month = jul,
year = {2020},
pages = {775--779},
}

@article{yang_deterministic_2024,
  title = {Deterministic Generation of Frequency-Bin-Encoded Microwave Photons},
  author = {Yang, Jiaying and Khanahmadi, Maryam and Strandberg, Ingrid and Gaikwad, Akshay and Castillo-Moreno, Claudia and Kockum, Anton Frisk and Ullah, Muhammad Asad and Johansson, G\"oran and Eriksson, Axel Martin and Gasparinetti, Simone},
  journal = {Phys. Rev. Lett.},
  volume = {134},
  issue = {24},
  pages = {240803},
  numpages = {9},
  year = {2025},
  month = {Jun},
  publisher = {American Physical Society},
  doi = {10.1103/PhysRevLett.134.240803},
  url = {https://link.aps.org/doi/10.1103/PhysRevLett.134.240803}
}

@article{qiu_deterministic_2025,
title = {Deterministic quantum state and gate teleportation between distant superconducting chips},
volume = {70},
issn = {20959273},
url = {https://linkinghub.elsevier.com/retrieve/pii/S209592732400879X},
doi = {10.1016/j.scib.2024.11.047},
number = {3},
urldate = {2025-03-05},
journal = {Science Bulletin},
author = {Qiu, Jiawei and Liu, Yang and Hu, Ling and Wu, Yukai and Niu, Jingjing and Zhang, Libo and Huang, Wenhui and Chen, Yuanzhen and Li, Jian and Liu, Song and Zhong, Youpeng and Duan, Luming and Yu, Dapeng},
month = feb,
year = {2025},
pages = {351--358},
}

@article{mcintyre_protocols_2025,
  title = {Protocols for intermodule two-qubit gates mediated by time-bin encoded photons},
  author = {McIntyre, Z. M. and Coish, W. A.},
  journal = {Phys. Rev. Res.},
  volume = {7},
  issue = {2},
  pages = {023255},
  numpages = {16},
  year = {2025},
  month = {Jun},
  publisher = {American Physical Society},
  doi = {10.1103/xwyt-1ck7},
  url = {https://link.aps.org/doi/10.1103/xwyt-1ck7}
}

@article{miyamura_generation_2025,
  title = {Generation of Frequency-Tunable Shaped Single Microwave Photons Using a Fixed-Frequency Superconducting Qubit},
  author = {Miyamura, Takeaki and Sunada, Yoshiki and Wang, Zhiling and Ilves, Jesper and Matsuura, Kohei and Nakamura, Yasunobu},
  journal = {PRX Quantum},
  volume = {6},
  issue = {2},
  pages = {020347},
  numpages = {16},
  year = {2025},
  month = {Jun},
  publisher = {American Physical Society},
  doi = {10.1103/PRXQuantum.6.020347},
  url = {https://link.aps.org/doi/10.1103/PRXQuantum.6.020347}}

@article{almanakly_deterministic_2025,
title = {Deterministic remote entanglement using a chiral quantum interconnect},
copyright = {2025 The Author(s), under exclusive licence to Springer Nature Limited},
issn = {1745-2481},
url = {https://www.nature.com/articles/s41567-025-02811-1},
doi = {10.1038/s41567-025-02811-1},
urldate = {2025-04-01},
volume = {21},
journal = {Nature Phys},
author = {Almanakly, Aziza and Yankelevich, Beatriz and Hays, Max and Kannan, Bharath and Assouly, Réouven and Greene, Alex and Gingras, Michael and Niedzielski, Bethany M. and Stickler, Hannah and Schwartz, Mollie E. and Serniak, Kyle and Wang, Joel I.-J. and Orlando, Terry P. and Gustavsson, Simon and Grover, Jeffrey A. and Oliver, William D.},
month = mar,
year = {2025},
pages = {825--830},
}

@article{caleffi_distributed_2024,
title = {Distributed quantum computing: {A} survey},
volume = {254},
issn = {13891286},
shorttitle = {Distributed quantum computing},
url = {https://linkinghub.elsevier.com/retrieve/pii/S1389128624005048},
doi = {10.1016/j.comnet.2024.110672},
urldate = {2025-04-05},
journal = {Computer Networks},
author = {Caleffi, Marcello and Amoretti, Michele and Ferrari, Davide and Illiano, Jessica and Manzalini, Antonio and Cacciapuoti, Angela Sara},
month = dec,
year = {2024},
pages = {110672},
}

@article{van_damme_advanced_2024,
title = {Advanced {CMOS} manufacturing of superconducting qubits on 300 mm wafers},
volume = {634},
copyright = {2024 The Author(s)},
issn = {1476-4687},
url = {https://www.nature.com/articles/s41586-024-07941-9},
doi = {10.1038/s41586-024-07941-9},
number = {8032},
urldate = {2025-04-05},
journal = {Nature},
author = {Van Damme, J. and Massar, S. and Acharya, R. and Ivanov, Ts and Perez Lozano, D. and Canvel, Y. and Demarets, M. and Vangoidsenhoven, D. and Hermans, Y. and Lai, J. G. and Vadiraj, A. M. and Mongillo, M. and Wan, D. and De Boeck, J. and Potočnik, A. and De Greve, K.},
month = oct,
year = {2024},
pages = {74--79},
}

@article{osman_mitigation_2023,
title = {Mitigation of frequency collisions in superconducting quantum processors},
volume = {5},
issn = {2643-1564},
url = {https://link.aps.org/doi/10.1103/PhysRevResearch.5.043001},
doi = {10.1103/PhysRevResearch.5.043001},
number = {4},
urldate = {2025-04-05},
journal = {Phys. Rev. Research},
author = {Osman, Amr and Fernández-Pendás, Jorge and Warren, Christopher and Kosen, Sandoko and Scigliuzzo, Marco and Frisk Kockum, Anton and Tancredi, Giovanna and Fadavi Roudsari, Anita and Bylander, Jonas},
month = oct,
year = {2023},
pages = {043001},
}

@article{PMagnard_2020,
  title = {Microwave Quantum Link between Superconducting Circuits Housed in Spatially Separated Cryogenic Systems},
  author = {Magnard, P. and Storz, S. and Kurpiers, P. and Sch\"ar, J. and Marxer, F. and L\"utolf, J. and Walter, T. and Besse, J.-C. and Gabureac, M. and Reuer, K. and Akin, A. and Royer, B. and Blais, A. and Wallraff, A.},
  journal = {Phys. Rev. Lett.},
  volume = {125},
  issue = {26},
  pages = {260502},
  numpages = {7},
  year = {2020},
  month = {Dec},
  publisher = {American Physical Society},
  doi = {10.1103/PhysRevLett.125.260502},
  url = {https://link.aps.org/doi/10.1103/PhysRevLett.125.260502}
}

@misc{gandotra_2025,
      title={Quantum communication over bandwidth-and-time-limited channels}, 
      author={Aditya Gandotra and Zhaoyou Wang and Aashish A. Clerk and Liang Jiang},
      year={2025},
      eprint={2502.08831},
      archivePrefix={arXiv},
      primaryClass={quant-ph},
      url={https://arxiv.org/abs/2502.08831}, 
}

@article{yamamoto_JPA_2008,
      title={Flux-driven {Josephson} parametric amplifier}, 
      author={T. Yamamoto and K. Inomata and M. Watanabe and K. Matsuba and T. Miyazaki and W. D. Oliver and Y. Nakamura and J. S. Tsai},
      journal = {Appl. Phys. Lett.},
      volume = {93},
      issue = {4},
      pages = {042510},
      year = {2008},
      month = {July},
      doi={ https://doi.org/10.1063/1.2964182}, 
}

@article{Sete_bandpass_purcell_2015,
  title = {Quantum theory of a bandpass {Purcell} filter for qubit readout},
  author = {Sete, Eyob A. and Martinis, John M. and Korotkov, Alexander N.},
  journal = {Phys. Rev. A},
  volume = {92},
  issue = {1},
  pages = {012325},
  numpages = {13},
  year = {2015},
  month = {Jul},
  publisher = {American Physical Society},
  doi = {10.1103/PhysRevA.92.012325},
  url = {https://link.aps.org/doi/10.1103/PhysRevA.92.012325}
}

@article{Jeffrey_bandpass_purcell_2014,
  title = {Fast Accurate State Measurement with Superconducting Qubits},
  author = {Jeffrey, Evan and Sank, Daniel and Mutus, J. Y. and White, T. C. and Kelly, J. and Barends, R. and Chen, Y. and Chen, Z. and Chiaro, B. and Dunsworth, A. and Megrant, A. and O'Malley, P. J. J. and Neill, C. and Roushan, P. and Vainsencher, A. and Wenner, J. and Cleland, A. N. and Martinis, John M.},
  journal = {Phys. Rev. Lett.},
  volume = {112},
  issue = {19},
  pages = {190504},
  numpages = {5},
  year = {2014},
  month = {May},
  publisher = {American Physical Society},
  doi = {10.1103/PhysRevLett.112.190504},
  url = {https://link.aps.org/doi/10.1103/PhysRevLett.112.190504}
}

@article{Sunada_intrinsic_purcell_2022,
  title = {Fast Readout and Reset of a Superconducting Qubit Coupled to a Resonator with an Intrinsic {Purcell} Filter},
  author = {Sunada, Y. and Kono, S. and Ilves, J. and Tamate, S. and Sugiyama, T. and Tabuchi, Y. and Nakamura, Y.},
  journal = {Phys. Rev. Appl.},
  volume = {17},
  issue = {4},
  pages = {044016},
  numpages = {12},
  year = {2022},
  month = {Apr},
  publisher = {American Physical Society},
  doi = {10.1103/PhysRevApplied.17.044016},
  url = {https://link.aps.org/doi/10.1103/PhysRevApplied.17.044016}
}

@article{Spring_intrinsic_purcell_2025,
  title = {Fast Multiplexed Superconducting-Qubit Readout with Intrinsic {Purcell} Filtering Using a Multiconductor Transmission Line},
  author = {Spring, Peter A. and Milanovic, Luka and Sunada, Yoshiki and Wang, Shiyu and van Loo, Arjan F. and Tamate, Shuhei and Nakamura, Yasunobu},
  journal = {PRX Quantum},
  volume = {6},
  issue = {2},
  pages = {020345},
  numpages = {23},
  year = {2025},
  month = {Jun},
  publisher = {American Physical Society},
  doi = {10.1103/PRXQuantum.6.020345},
  url = {https://link.aps.org/doi/10.1103/PRXQuantum.6.020345}
}

@article{Raymer_TM_review_2020,
  title = {Temporal modes in quantum optics: then and now},
  author = {Michael G Raymer and Ian A Walmsley},
  journal = {Phys. Scr.},
  volume = {95},
  number = {6},
  pages = {064002},
  year = {2020},
  month = {Mar},
  publisher = {American Physical Society},
  doi = {10.1088/1402-4896/ab6153},
  url = {https://iopscience.iop.org/article/10.1088/1402-4896/ab6153}
}

@article{Reddy_QPG_2018,
  title = {High-selectivity quantum pulse gating of photonic temporal modes using all-optical Ramsey interferometry},
  author = {Dileep V. Reddy and Michael G. Raymer},
  journal = {Optica},
  volume = {5},
  pages = {423--428},
  year = {2018},
  doi = {10.1364/OPTICA.5.000423},
  url = {https://opg.optica.org/optica/fulltext.cfm?uri=optica-5-4-423&id=385263}
}

@article{SKrinner_heatload_2019,
  title = {Engineering cryogenic setups for 100-qubit scale superconducting circuit systems},
  author = {Krinner, Sebastian and Storz, Simon and Kurpiers, Philipp and Magnard, Paul and Heinsoo, Johannes and Keller, Raphael and Lütolf, Janis and Eichler, Christopher and Wallraff, Andreas},
  journal = {EPJ Quantum Technology},
  volume = {6},
  pages = {2},
  year = {2019},
  doi = {10.1140/epjqt/s40507-019-0072-0},
  url = {https://doi.org/10.1140/epjqt/s40507-019-0072-0}
}

@article{JICirac_commtheory_1997,
  title = {Quantum State Transfer and Entanglement Distribution among Distant Nodes in a Quantum Network},
  author = {Cirac, J. I. and Zoller, P. and Kimble, H. J. and Mabuchi, H.},
  journal = {Phys. Rev. Lett.},
  volume = {78},
  issue = {16},
  pages = {3221--3224},
  numpages = {0},
  year = {1997},
  month = {Apr},
  publisher = {American Physical Society},
  doi = {10.1103/PhysRevLett.78.3221},
  url = {https://link.aps.org/doi/10.1103/PhysRevLett.78.3221}
}

@article{ZWang_freq-bin_2025,
  title = {Generation of Frequency-Bin-Encoded Dual-Rail Cluster States via Time-Frequency Multiplexing of Microwave Photonic Qubits},
  author = {Wang, Zhiling and Miyamura, Takeaki and Sunada, Yoshiki and Sunada, Keika and Ilves, Jesper and Matsuura, Kohei and Nakamura, Yasunobu},
  journal = {PRX Quantum},
  volume = {7},
  issue = {1},
  pages = {010330},
  numpages = {21},
  year = {2026},
  month = {Feb},
  publisher = {American Physical Society},
  doi = {10.1103/rrct-dpfv},
  url = {https://link.aps.org/doi/10.1103/rrct-dpfv}
}

@article{JOsullivan_frequebcy-bin_2025,
title = {Deterministic generation of two-dimensional multi-photon cluster states},
volume = {16},
url = {https://doi.org/10.1038/s41467-025-60472-3},
doi = {10.1038/s41467-025-60472-3},
number = {5505},
journal = {Nature Communications},
author = {
    James O'Sullivan and
    Kevin Reuer and
    Aleksandr Grigorev and
    Xi Dai and
    Alonso Hernández-Antón and
    Manuel H. Muñoz-Arias and
    Christoph Hellings and
    Alexander Flasby and
    Dante Colao Zanuz and
    Jean-Claude Besse and
    Alexandre Blais and
    Daniel Malz and
    Christopher Eichler and
    Andreas Wallraff
},
month = {Jul},
year = {2025}
}

@article{RBianchetti_QPT_2010,
  title = {Control and Tomography of a Three Level Superconducting Artificial Atom},
  author = {Bianchetti, R. and Filipp, S. and Baur, M. and Fink, J. M. and Lang, C. and Steffen, L. and Boissonneault, M. and Blais, A. and Wallraff, A.},
  journal = {Phys. Rev. Lett.},
  volume = {105},
  issue = {22},
  pages = {223601},
  numpages = {4},
  year = {2010},
  month = {Nov},
  publisher = {American Physical Society},
  doi = {10.1103/PhysRevLett.105.223601},
  url = {https://link.aps.org/doi/10.1103/PhysRevLett.105.223601}
}

@misc{TMiyamura_qcomm_2025,
      title={Deterministic Quantum Communication Between Fixed-Frequency Superconducting Qubits via Broadband Resonators}, 
      author={Takeaki Miyamura and Zhiling Wang and Kohei Matsuura and Yoshiki Sunada and Keika Sunada and Kenshi Yuki and Jesper Ilves and Yasunobu Nakamura},
      year={2025},
      eprint={2512.08328},
      archivePrefix={arXiv},
      primaryClass={quant-ph},
      url={https://arxiv.org/abs/2512.08328}, 
}

@article{AHKiilerich_vc_2020,
  title = {Quantum interactions with pulses of radiation},
  author = {Kiilerich, Alexander Holm and M\o{}lmer, Klaus},
  journal = {Phys. Rev. A},
  volume = {102},
  issue = {2},
  pages = {023717},
  numpages = {14},
  year = {2020},
  month = {Aug},
  publisher = {American Physical Society},
  doi = {10.1103/PhysRevA.102.023717},
  url = {https://link.aps.org/doi/10.1103/PhysRevA.102.023717}
}

@article{Paye_chronocyclic_Wigner_1992,
    title={The chronocyclic representation of ultrashort light pulses},
    volume={28},
    url={http://dx.doi.org/10.1109/3.159533},
    doi={10.1109/3.159533},
    number={10},
    journal={IEEE Journal of Quantum Electronics},
    publisher={Institute of Electrical and Electronics Engineers (IEEE)},
    author={Paye, J.},
    year={1992},
    pages={2262–2273} }

@article{YSunada_phase-insensitive_2024,
  title = {Efficient Tomography of Microwave Photonic Cluster States},
  author = {Sunada, Yoshiki and Kono, Shingo and Ilves, Jesper and Sugiyama, Takanori and Suzuki, Yasunari and Okubo, Tsuyoshi and Tamate, Shuhei and Tabuchi, Yutaka and Nakamura, Yasunobu},
  journal = {PRX Quantum},
  volume = {7},
  issue = {1},
  pages = {010323},
  numpages = {28},
  year = {2026},
  month = {Feb},
  publisher = {American Physical Society},
  doi = {10.1103/3vnw-yjd5},
  url = {https://link.aps.org/doi/10.1103/3vnw-yjd5}
}

\end{document}